\documentclass[11pt,letterpaper]{article}
\usepackage{latexsym}
\usepackage{amssymb}
\usepackage{amsmath}
\usepackage{graphicx}
\newcommand{\hamma}{\gamma}

\begin{document}
\title{\bf{\Large Phase-space and Black Hole Entropy of Higher Genus Horizons in Loop Quantum Gravity}}
\author {{\small S. Kloster \footnote{stevek@sfu.ca}} \\
\it{\small Centre for Experimental and Constructive Mathematics,  Simon Fraser University} \\
\it{\small Burnaby, British Columbia, V5A 1S6, Canada}
 \and
{\small J. Brannlund \footnote{johanb@mathstat.dal.ca}} \\
\it{\small Department of Mathematics and Statistics, Dalhousie University} \\
\it{\small Halifax, Nova Scotia, B3H 
3J5, Canada }
\and
{\small A. DeBenedictis \footnote{adebened@sfu.ca}} \\
\it{\small Pacific Institute for the Mathematical Sciences,} \\
\it{\small Simon Fraser University Site} \\
\it{\small and}\\
\it{\small Department of Physics, Simon Fraser University}\\
\it{\small Burnaby, British Columbia, V5A 1S6, Canada }
}
\date{{\small February 20, 2008}}
\maketitle

\begin{abstract}
\noindent In the context of loop quantum gravity, we construct the
phase-space of isolated horizons with genus greater than $0$.
Within the loop quantum gravity framework, these horizons are
described by genus $g$ surfaces with $N$ punctures and the dimension of the
corresponding phase-space is calculated including the genus
cycles as degrees of freedom. From this, the black hole entropy
can be calculated by counting the microstates which correspond to
a black hole of fixed area. We find that the leading term agrees
with the $A/4$ law and that the sub-leading contribution is
modified by the genus cycles.
\end{abstract}

\vspace{3mm}

\section{Introduction}
The source of the apparent entropy of black holes has been a
fascinating area of study since Bekenstein's original calculation
\cite{ref:bek}. It is now well known that, to leading order, this
entropy has a value equal to one-quarter of the horizon area (in
proper units). Since the original calculation, many methods have
been employed in order to calculate this entropy (see
\cite{ref:entreview}, \cite{ref:entreview2} and references therein
for excellent reviews of the subject).

There is still debate on the actual source of this entropy. One
belief is that the source is strictly gravitational in origin.
That is, one should be able to define microstates in a full
quantum theory of gravity which, when counted, should yield the
correct entropy law. One promising approach to a theory of
quantum gravity is loop quantum gravity. This theory is
essentially a theory of quantum Riemannian geometry and seems to
reconcile principles of quantum mechanics with those of general
relativity. The subject has matured over the years and now there
exist many excellent reviews on the subject \cite{ref:thiem},
\cite{ref:rovbook}, \cite{ref:rev1}, \cite{ref:rev2}, including
several specifically related to black holes \cite{ref:bojo},
\cite{ref:modesto}. For studies directly related to the problem
of black hole entropy, the reader is referred to
\cite{ref:entrev1}, \cite{ref:ashbaez}, \cite{ref:daskaul}, \cite{ref:chatmaj}, \cite{ref:domlew},
\cite{ref:entrev2}, \cite{ref:entrev3}, \cite{ref:entrev3_5}, \cite{ref:dms}, 
\cite{ref:entrev4}, \cite{ref:entrev5}, \cite{ref:entrev6}.

\begin{figure}[ht]
\begin{center}
\includegraphics[bb=0 0 574 390, clip, scale=0.5, keepaspectratio=true]{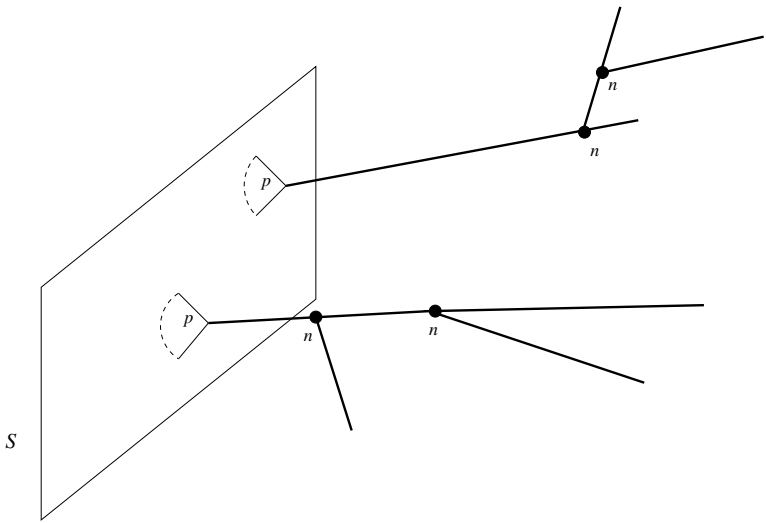}
\caption{{\small A gravitational spin-network endowing a surface
$S$ with geometry. The punctures $p$ can be pictured as
``tugging'' on the surface, endowing it with geometry and
introducing a local angular defect on the surface. The $n$'s are
the nodes, associated with volumes.}} \label{fig:0}
\end{center}
\end{figure}

In loop quantum gravity, a theory of $SU(2)$ spin-networks is
employed. The nodes of these networks are associated with quantum
volumes and the punctures that the networks make with a
surface endows it with an area, by introducing an angular defect
on the surface (see figure \ref{fig:0}). In the canonical
framework an ADM decomposition of the full spacetime is first
carried out and the relevant quantities are constructed on the ADM
three-surfaces.

The relevant canonically conjugate variables in loop quantum
gravity are the gravitational spin-connection, $A^{k}_{a}$ and a
densitized triad, $E^{a}_{k}$. Here, the
indices $a,\,b,\,c$ etc. denote the coordinates on the
three-surface and indices $k,\,l,\,m$ etc. denote $su(2)$ internal indices.

The relation of these operators to ``classical'' geometric
objects such as areas is quite intriguing. The operator
corresponding to areas is given by
\begin{equation}
 \hat{A}(S)=\int_{S}\sqrt{n_{a}\hat{E}^{a}_{k}\,n_{b}\hat{E}^{b}_{k}}\,d\sigma\: \label{eq:area}
\end{equation}
and has the following spectrum:
\begin{equation}
 \hat{A}(S)\left|\mathbf{S}\right> =8\pi\gamma\sum_{p} \sqrt{j_{p}(j_{p}+1)}
 \left|\mathbf{S}\right>, \label{eq:aevals}
\end{equation}
where $S$ denotes some surface. $p$ denotes which puncture is
under consideration and $j_{p}$ can take on half-integer values which represent spins carried by the punctures.
The normals to the surface are denoted by $n_{a}$ and $n_{b}$.
$\gamma$ is the Immirzi parameter, which is to be determined by
some means such as black hole entropy calculations \cite{ref:entrev2}, \cite{ref:entrev3},
\cite{ref:entrev6}, \cite{ref:immirz1}, \cite{ref:jacobson}. The assumption in the
above formula is that there are no spin network nodes on $S$ nor
any components of the spin network that are tangential to $S$. In
general, when considering a surface, the spin network can go
``straight through'' the surface or it can bifurcate on the
surface. However, the contribution from the latter is negligible
for large surfaces. Also, it has been shown that acceptable
quantum states in horizon entropy calculations are only those
which count punctures that go straight-through the surface and
are stable to small deformations of the surface \cite{ref:kras}.
Therefore, the areas given by (\ref{eq:aevals}) are those
relevant for the calculation here. A nice explanation of this can
also be found in \cite{ref:entrev5}. The consequences of relaxing this counting condition, and taking into account bifurcations on the horizon, is presented in \cite{ref:countambig}.

In the above references, the situation for spherical horizons has
been studied in depth. However, General Relativity allows for the
existence of horizons of other geometries and topologies.
Interestingly, if one admits a negative cosmological constant
into the theory, solutions with cylindrical, toroidal and higher genus topology
are also admitted \cite{ref:rg1}, \cite{ref:tor1}, \cite{ref:tor2}, \cite{ref:vanzo},
\cite{ref:tor3}, \cite{ref:rg2}, \cite{ref:rg3}, \cite{ref:tor4}, \cite{ref:liko}, \cite{ref:mena}). Several of these papers have studied the entropy issue from a non loop quantum gravity perspective (\cite{ref:rg2} and \cite{ref:rg3} consider dilatonic and dimensionally continued black holes). Granted, these solutions are
not generally considered to be of astrophysical interest at least
due to the fact that observations seem to favor a deSitter
universe. However, the anti-deSitter solutions with exotic
topology have been studied in detail as they provide a rich arena
in which to check internal consistencies of general relativity
theory. In this vein these types of black holes are also
important in studies of black hole entropy using thermodynamic and
quantum field theoretic techniques \cite{ref:thermo}. This is what
motivates us to study such black holes in the paradigm of loop
quantum gravity. 

In loop quantum gravity the entropy arises from counting the number of ways spin-networks can puncture the horizon surface and endow it with a fixed net area. That is, the number of quantum states corresponding to the same area are counted. This is a well defined quantity and therefore, in this paradigm, there is no problem in assigning a quantum statistical entropy to black holes even in asyptotically anti-deSitter space-time.

One may find many forms of metrics describing higher genus black holes in anti-de Sitter space-time. We will take a sufficiently general form which can be written as:
\begin{equation}
ds^2=-\left(\alpha^2 \rho^2 -b -\frac{2M}{\rho}\right)\,dt^{2} + \frac{d\rho^{2}}{\left(\alpha^2 \rho^2 -b -\frac{2M}{\rho}\right)} +\rho^2\left(d\theta^{2} + a\sinh^{2}(\sqrt{b}\,\theta)\,d\phi^{2}\right)\:, \label{eq:smmetric}
\end{equation}
with $0 < \phi \leq 2\pi$. Here, $\alpha$ is related to the cosmological constant via $\alpha^{2}=-\Lambda/3$, $M$ is the mass parameter, and $a$ and $b$ are constants that determine the topology of $t,\rho=$constant surfaces. The cases are as follows:\\
i) $b=-1$, $a=-1$: In this case constant $(t,\,\rho)$ surfaces are spheres. We will not consider this case as it was studied in \cite{ref:ashbaez}. \\
ii) $b=0$, $\underset{b\rightarrow 0}\lim\,a=\frac{1}{b}$: In this case constant $(t,\,\rho)$ surfaces are tori. We will consider this case separately from the higher genus case. \\
iii) $b=1$, $a=1$: In this case constant $(t,\,\rho)$ surfaces are surfaces of constant negative curvature of genus $g > 1$, depending on the identifications chosen. \\
An event horizon exists when $\left(\alpha^2 \rho^2 -b -\frac{2M}{\rho}\right)=0$.

At the horizon, isolated horizon boundary conditions are employed. Isolated horizons with negative cosmological constant, mainly with $S^{2}$ topology, are defined
in detail in \cite{ref:ashlambda}. Some other interesting studies
of isolated horizons, without cosmological constant include
\cite{ref:ashbaez}, \cite{ref:ashcor}, \cite{ref:ashfair},
\cite{ref:ashengle}, \cite{ref:ashengle2}. We generalize the situation here to accommodate higher genus horizons, which is reasonably straight-forward and is mentioned in \cite{ref:abf}.

It can be noted that the black hole metrics given by (\ref{eq:smmetric}) possess all the relevant features of isolated horizons needed here. This can most easily be checked by calculating the Newman-Penrose spin coefficients and comparing them with the isolated horizon requirements outlined in \cite{ref:ashcor} and \cite{ref:jerzlewandhoriz}. We next summarize the main steps leading to the symplectic structure.

In the calculations that follow, explicit properties of the metrics (\ref{eq:smmetric}) are not utilized and therefore the properties of these geometries are not a priori assumed. Only the fact that they are isolated horizons is used. However, the metrics (\ref{eq:smmetric}) can serve to provide a valuable consistency check. 

At the isolated horizon the canonical variables do not decouple as they do at infinity \cite{ref:ashcor}. Therefore an extra term is present in the action introduced by the variation with respect to the connection (indices suppressed):
\begin{equation}
 \delta S_{|\Delta}= -\frac{i}{8\pi}\int_{\Delta} \mbox{Tr}\left[\Sigma \wedge \delta A\right]\: . \label{eq:deltavar}
\end{equation}
where $\Delta$ is the null submanifold corresponding to the horizon. The two-form $\Sigma$ is related to the triad via $E^{a}_{\;i}:=\gamma \eta^{abc} \Sigma_{bci}$, with $\eta^{abc}$ the Levi-Civita density of weight 1. 

The curvature of the connection is given by $F=dA+A\wedge A$. The field equations at $\Delta$ allow us to express the pull back of $F$ (pulled back to $\Delta$ and denoted by an under-arrow) in terms of the pull-back of the Riemann tensor via \footnote{The indices here differ from the earlier expression, reflecting a different representation of the algebra. The relation between the two is $\Sigma_{abA}^{\;\;\;B}=-\frac{i}{2}\Sigma^{i}_{\;ab} \tau_{iA}^{\;\;B}$.}
\begin{equation}
\underleftarrow{F}_{ab}^{\;\;AB}= -\frac{1}{4}\underleftarrow{R}_{ab}^{\;\;cd}\,\Sigma^{AB}_{cd} \:, \label{riemtosigma}
\end{equation}
where it is implied that this equation holds at $\Delta$. 

In brief, the properties of the Weyl scalars (customarily denoted by $\Psi$) as well as the Ricci spinor components (denoted by $\Phi$)  allow us to express the pull-back of $F$ as \cite{ref:ashcor}:
\begin{equation}
 \underleftarrow{F}_{ab}^{\;\;AB}=\left[ \left(\Psi_{2} - \Phi_{11} - \frac{R}{24}\right)\delta^{A}_{\;C} \delta^{B}_{\;D}  -\left(\frac{3}{2}\Psi_{2} - \Phi_{11}\right) o^{A}o^{B} i_{C}i_{D}\right]  \underleftarrow{\Sigma}_{ab}^{\;\;CD}\:, \label{eq:pullback}
\end{equation}
where we are using the standard notation convention ($i_{A}$ and $o_{A}$) for the spinor dyad (see \cite{ref:ashcor} and \cite{ref:chandra} for details). The  Ricci spinor component, $\Phi_{11}$, although zero in the case  where there is no energy flux nor shear stresses present, and the bulk equations of motion hold, is left here for generality so that no assumptions are made about the components at the moment, save that they satisfy the conditions for an isolated horizon.

Before proceeding, we note that the boundary conditions on the isolated horizon imply that $\delta A_{a}^{AB} o_{A}o_{B} = 0$ \cite{ref:ashcor}. Therefore, we can re-write (\ref{eq:deltavar}), noting that the integrand is pulled back to $\Delta$, as:
\begin{equation}
 \delta S_{|\Delta}= -\frac{i}{8\pi}\int_{\Delta} \mbox{Tr}\left[\frac{1}{\left(\Psi_{2} - \Phi_{11} -\frac{R}{24}\right)} \underleftarrow{F}\wedge \delta \underleftarrow{A}\right]\:.
\end{equation}
The quantity in round parentheses is constant on the horizon \footnote{In the language often found in the literature on isolated horizons, it is ``spherically symmetric''. However, it should be noted that this does \emph{not} imply that the surface $\Delta$ must be a sphere.}. Furthermore, it can easily be checked with metric (\ref{eq:smmetric}) that it equals $-\frac{1}{4} ^{(2)}R$, with $^{(2)}R$ being the two-dimensional Ricci scalar on the horizon. One can then, by use of the Gauss-Bonnet theorem, deduce that:
\begin{equation}
\left(\Psi_{2} - \Phi_{11}- \frac{R}{24}\right)=-\frac{2\pi (1-g)}{a_{0}}, \label{eq:coefficient}
\end{equation}
with $a_{0}$ the (fixed) area of the horizon. (One can also verify this by explicit calculation with the metric (\ref{eq:smmetric})).  It is therefore concluded that
\begin{equation}
 \delta S_{|\Delta}= \frac{i}{8\pi}\frac{a_{0}}{2\pi (1-g)}\int_{\Delta} \mbox{Tr}\left[\underleftarrow{F}\wedge \delta \underleftarrow{A}\right]\:. \label{newdeltaform}
\end{equation}
At this point, the derivation follows that of \cite{ref:ashcor}, which is applicable in the higher genus case as well, with minor modification. It was shown in \cite{ref:ashcor} that the isolated horizon boundary conditions reduce the number degrees of freedom. The boundary theory actually reduces to a $U(1)$ theory and is of the form (generalised to arbitrary genus)
\begin{equation}
S_{|\Delta}= \frac{i}{8\pi}\frac{a_{0}}{2\pi (1-g)}\int_{\Delta} W \wedge dW \:, \label{eq:u1form}
\end{equation}
with $W$ a $U(1)$ connection on the boundary surface, which is restricted by
the value of the bulk $SU(2)$ connection penetrating the surface at
that particular point.

The procedure to deduce the symplectic form gives rise to the following symplectic
structure \cite{ref:ashbaez}, \cite{ref:ashcor}:
\begin{equation}
 \Omega_{grav}(.., \, ..)=\frac{1}{8\pi}\left[\int_{M} \mbox{Tr}\left(\delta A \wedge
  \delta^{\prime}\Sigma - \delta^{\prime}A \wedge \delta\Sigma\right)
  +4k \oint_{\partial M} \delta W \wedge \delta^{\prime}W\right], \label{eq:gravsymp}
\end{equation}
where $W$ and $W^{\prime}$ are $U(1)$ connections on the boundary
surface and $M$ is the bulk manifold. In essence, $W$
can be thought of as a mapping from a set of paths on the boundary of
$M$ into $U(1)$. The number of degrees of freedom in $W$ is the number
of values of $W$ we can assign before $W$ is completely fixed. $\delta W$
and $\delta^{\prime}W$ are tangent vectors in the space of $U(1)$
connections.

As noted in \cite{ref:ashbaez} and \cite{ref:ashcor}, the surface
term has the form of a Chern-Simons theory where the quantity $k=\frac{a_{0}}{4\pi(g-1)\gamma}$
is the Chern level of the boundary theory (a natural number). We
will construct the quantities in this surface term explicitly in
the next section. The case of $g=1$ is special but can be accommodated within this framework. This case will be discussed separately after the $g>1$ scenario.

The non-zero genus situations are quite interesting as one can include, in
addition to the degrees of freedom introduced by spin-network
punctures, degrees of freedom associated with the genus
cycles. For the BTZ black hole it has been shown, using a
semi-classical Euclidean path integral approach, that corrections
to the entropy arise from the toroidal boundary of the space-time
\cite{ref:govind}. Vanzo \cite{ref:vanzo}, Mann and Solodukhin\cite{ref:mans} and Liko \cite{ref:liko} also find dependence on topology in the black hole entropy utilizing non quantum gravity techniques. As well, studies of lower dimensional systems
often utilize toroidal cycles as relevant degrees of freedom
\cite{ref:lowdim1}, \cite{ref:lowdim2}. We show below that there is a natural way to introduce genus effects within the paradigm of loop quantum gravity.

In the following section we review the symplectic structure in the
case of the spherical boundary and construct, in detail, the
symplectic structure for toroidal and cylindrical horizons punctured by a
gravitational spin-network. The toroidal and cylindrical degrees of freedom are
included producing some interesting results. We relate this
construction to the number of degrees of freedom of the system
which, in turn, is directly related to the entropy as summarized
in section 3.

\section{The phase-space}
\subsection{A brief review of the spherical horizon}
\begin{figure}[ht]
\begin{center}
\includegraphics[bb=0 0 235 115, clip, scale=0.7, keepaspectratio=true]{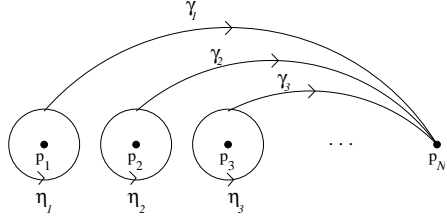}
\caption{{\small A set of paths used to define the symplectic
structure on an $N$-punctured sphere. The $\eta$ cycles encircle
the punctures whereas the $\gamma$ paths connect the punctures to
a ``base''-puncture, $p_{N}$.}} \label{fig:sphere}
\end{center}
\end{figure}

Before continuing we briefly review here the case of a spherical
horizon, which was pioneered and studied in detail in
 \cite{ref:ashbaez}, \cite{ref:ashcor}. In the case of spherical
horizons, one has a sphere with $N$-punctures due to the gravitational spin-network. The first cohomology
group of the $N$-punctured sphere, denoted as $H^{1}(S-P_{N})$, is $(N-1)$-dimensional which is one less
than the number of punctures. ($N-1$) pairs of forms are defined
on the punctured sphere to yield the required symplectic
structure (see figure \ref{fig:sphere}, which is similar to the
figure originally produced in \cite{ref:ashbaez}). These forms
are constructed via their duality with chains on a punctured
sphere as depicted in figure \ref{fig:sphere}.

There exist $N-1\;$ $\eta$ paths and $N-1$ conjugate $\gamma$
paths on this sphere. A basis for all the paths based at $p_{N}$
is
\begin{equation}
 \left\{\gamma^{-1}_{1}\eta_{1}\gamma_{1},\, \gamma^{-1}_{2}\eta_{2}\gamma_{2},\,...,\,\gamma^{-1}_{N-1}\eta_{N-1}\gamma_{N-1}\right\}.
\end{equation}
At first sight there seems to be an asymmetry, due to the lack of
a path $\eta_{N}$. However, there exists a fundamental relation
\begin{equation}
 \eta_{1}\cdot\eta_{2}\cdot ... \cdot \eta_{N}=1, \label{eq:etarel}
\end{equation}
which is a mathematical relation indicating that a loop around all
punctures can be shrunk to a point on the sphere. Another way to
look at this relation is that a loop around all the $N-1$
punctures is equivalent to a loop around the $N$-th puncture but
in reverse. In other words, $\eta_{N}$ is expressible in terms of
the other $\eta$ paths.

The set of paths on the sphere may be decomposed into a set of
loops $\left\{\eta_{1},\,\eta_{2},\,...,\,\eta_{N-1}\right\}$ and
a set of ``translations''
$\left\{\gamma_{1},\,\gamma_{2},\,...,\,\gamma_{N-1}\right\}$. If
$W$ and $W^{\prime}$ agree on these two sets, they will agree on
all paths on the sphere. Thus we have $2(N-1)$ degrees of freedom
or twice $dim\left[H^{1}(S-P_{N})\right]$.

Quantum states $\psi_{a}$ are then obtained for
$a=(a_{1},..,a_{N-1})$ with $a_{i}\in\left\{1, .., k\right\}$
\cite{ref:ashbaez}. The integers $a_{i}$ play a role similar to
the magnetic quantum number in quantum mechanics. The condition
(\ref{eq:etarel}) gives rise to a constraint:
\begin{equation}
 a_{1}+...+a_{N-1}=-a_{N}\;. \label{eq:projconst}
\end{equation}
 This restriction is the quantum analogue of the Gauss-Bonnet theorem for a sphere.
 Note that one now has $N$ generators and one constraint, matching the dimension
 of the first cohomology group. Thus, for a spherical horizon, states can be
 labeled with $a=(a_{1},..,a_{N})$
 subject to constraint (\ref{eq:projconst}).
 We will find below that in the case of higher genus horizons these relations are changed.

\subsection{The $g>1$ horizon}
\begin{figure}[ht]
\begin{center}
\includegraphics[bb=0 0 810 433, clip, scale=0.47 , keepaspectratio=true]{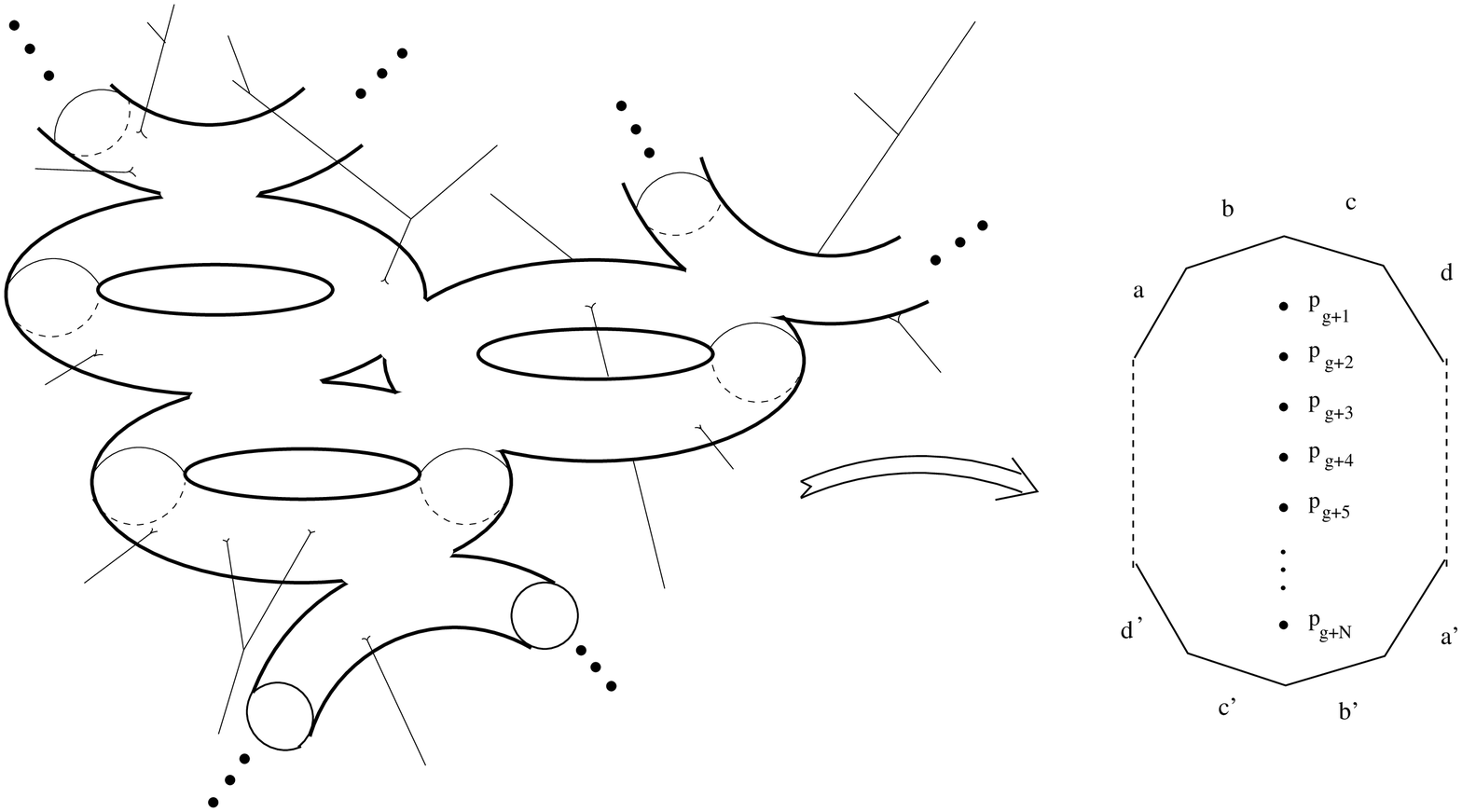}
\caption{{\small A qualitative pictorial representation of the horizon. This surface is punctured by the
gravitational spin network, giving it geometry and area. To
construct the phase-space we consider a polygon with 4$g$ sides and appropriate identification. The spin network punctures this polygon $N$ times.}} \label{genii}
\end{center}
\end{figure}

\begin{figure}[ht]
\begin{center}
\includegraphics[bb=0 0 1115 500, clip, scale=0.40, keepaspectratio=true]{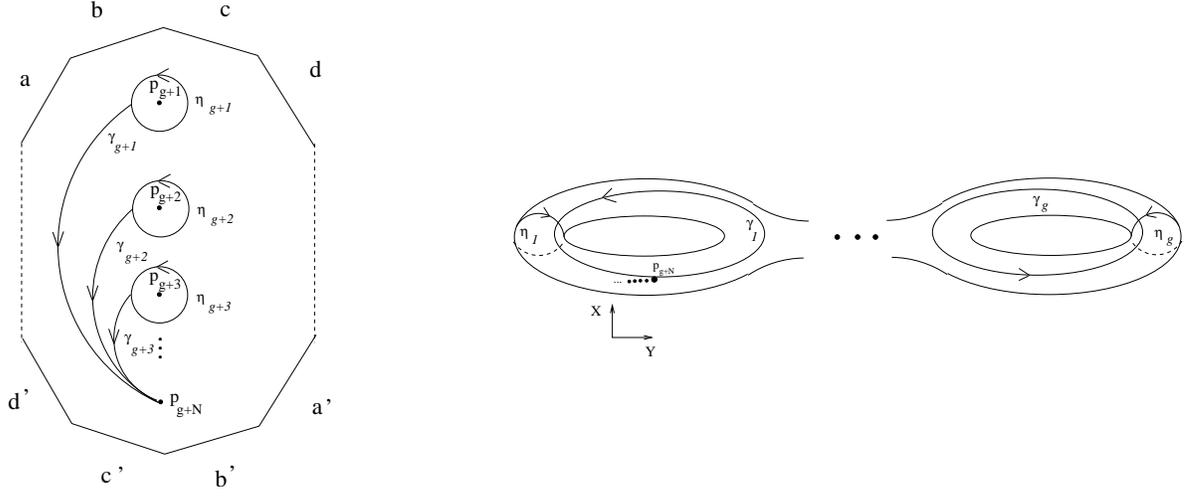}
\caption{{\small Open chains ($\gamma$'s) and cycles ($\eta$'s)
used to define forms for the symplectic structure of the
phase-space of the horizon. There are paths associated with the punctures (left) and the toroidal holes (right)}} \label{fig:cycles}
\end{center}
\end{figure}

To study the symplectic structure of the phase-space we must
construct sets of conjugate forms with appropriate properties to
yield a canonical symplectic form corresponding to the horizon in
question. To aid in constructing the symplectic structure on
the horizon, it is useful to picture the $t=$constant 2-surface of the black hole
horizon as a finite plane with 4$g$ sides with appropriate identification. The gravitational spin-network, which endows the surface
with geometry, punctures the surface $N$ times. The number of
punctures is sufficient to give the surface an area $A$ (see
figure \ref{genii}) according to (\ref{eq:aevals}).
As in figure \ref{fig:cycles}, pairs of open chains and cycles are
constructed on the punctured genus $g$ torus and their dual forms will be
utilized to construct the symplectic structure of the phase-space.

The chains and cycles in figure \ref{fig:cycles} are labeled as follows: The $\gamma$ and $\eta$ paths associated with the toroidal holes are denoted with subscripts $1$ to $g$. Those associated with the spin network punctures are denoted with subscripts $g+1$ to $g+N$. The paths associated with the $g+N$th puncture are not shown as this puncture will serve as the ``base'' point in our construction.

In the case of a $g > 1$ genus $g$ torus, consider a loop which goes once around the
edges of the identified polygon. This path is denoted as
$\eta_{1}\gamma_{1}\eta_{1}^{-1}\gamma_{1}^{-1} \cdot ... \cdot \eta_{g}\gamma_{g}\eta_{g}^{-1}\gamma_{g}^{-1}$ and is
equivalent to a loop going once around all the $N$ punctures,
\begin{equation}
 \eta_{g+1}\cdot\eta_{g+2}\cdot ... \cdot \eta_{g+N}= \eta_{1}\gamma_{1}\eta_{1}^{-1}\gamma_{1}^{-1} \cdot ... \cdot \eta_{g}\gamma_{g}\eta_{g}^{-1}\gamma_{g}^{-1}. \label{eq:torusrelg}
\end{equation}
This relation illustrates that
degrees of freedom can be attributed to the toroidal loops.

A basis for all paths based at $p_{g+N}$ will break down to
$\left\{\eta_{1},\,...,\,\eta_{g+N-1}\right\}$ and
$\left\{\gamma_{1},\,...,\,\gamma_{g+N-1}\right\}$. If
$W$ and $W^{\prime}$ agree on these two sets, they will agree on
all paths. Note that, like the spherical case, $\eta_{N}$ again
is not used as it can be expressed using relation
(\ref{eq:torusrelg}). However, note that the genus cycles, $\eta_{\leq g}$ \emph{are}
utilized and there is nothing to constrain the values assigned to the
$W(\eta_{\leq g})$. Hence,
$\eta_{\leq g}$ and $\gamma_{\leq g}$ contribute new degrees of freedom. The
result is $2(N+g-1)$ degrees of freedom and is thus not twice the
dimension of the first cohomology group on an $N$-punctured genus $g$ torus
(which is $dim\left[H^{1}(T_{g}-P_{N})\right]=N+2g-1$).

We denote the forms dual to the $\gamma$-paths as $\alpha$ and
the forms dual to the $\eta$-paths as $\beta$.

The dual forms satisfy the following properties:
\begin{enumerate}
 \item
 \begin{align}
  &\int_{\gamma_{j}}\,\alpha_{i}= \delta_{ij}\:,&\oint_{\eta_{j}}\,\alpha_{i}=0\:, \nonumber \\
  &\oint_{\eta_{n}}\,\alpha_{i}=0\:,&\int_{\gamma_{i}} \,\alpha_{m}=0\:,  \nonumber \\
  & \int_{\gamma_{n}} \,\alpha_{m}=\delta_{mn}\:,& \int_{\gamma_{n}}\,\alpha_{i}=0, \nonumber \\
  & \oint_{\eta_{n}}\,\alpha_{m}=0\:, & \oint_{\eta_{i}}\,\alpha_{m}=0. \nonumber
 \end{align}
  \item
  \begin{align}
  &\oint_{\eta_{j}}\,\beta_{i}= \delta_{ij}\:,&\int_{\gamma_{j}}\,\beta_{i}=0\:, \nonumber \\
  &\int_{\gamma_{n}}\,\beta_{i}=0\:,&\oint_{\eta_{i}} \,\beta_{m}=0\:,  \nonumber \\
  & \oint_{\eta_{n}} \,\beta_{m}=\delta_{mn}\:,& \oint_{\eta_{n}}\,\beta_{i}=0, \nonumber \\
  & \int_{\gamma_{n}}\,\beta_{m}=0\:, & \oint_{\gamma_{i}}\,\beta_{m}=0. \nonumber
 \end{align}
  \item
  \begin{align}
  &\int_{T}\,\alpha_{i}\wedge \alpha_{j}= 0\:,\;\;\; \int_{T}\,\beta_{i}\wedge
 \beta_{j}=0\:,\;\;\;
  \int_{T}\,\alpha_{i}\wedge \beta_{j}=\delta_{ij}\:, \nonumber
  \\
  &\int_{T}\,\alpha_{n}\wedge \beta_{m}=\delta_{mn}\:, \;\;\; \int_{T}\,\alpha_{n}\wedge\beta_{j}=0\:,\;\;\;
  \int_{T} \alpha_{i} \wedge \beta_{m}=0, \nonumber \\
  &\int_{T}\, \alpha_{i} \wedge \alpha_{m} =0\:,\;\;\; \int_{T}\, \beta_{i} \wedge \beta_{m} =0\:. \nonumber
 \end{align}
\end{enumerate}
Here $i,\,j=g+1,...\,,g+N-1$ are indices denoting chains and dual
forms associated with the $N-1$ punctures (not including the
``base'' puncture, $p_{g+N}$) on the surface (the surface being denoted as $T$) whereas the subscripts
$m$ and $n$ denote the chains and dual forms associated with the
toroidal holes (see figure \ref{fig:cycles}) and therefore take on values $1 \, ...\, g$. We also demand that the
$\beta_{i}$ possess simple singularities at the punctures and
that $W$ is flat everywhere, save for the punctures (see
\cite{ref:ashbaez} for details).

In this construction, chains in the direction of the genus $g$ torus angles
 are taken as conjugates of
each other, provided they are associated with the same genus hole. This is in much the same way as $\eta_{i}$ and $\gamma_{i}$
are conjugates. It turns out that this is the natural choice if
we wish to define a non-degenerate symplectic two-form on this surface.

Forms with some of the properties in 2 and 3 can be shown to
exist from the deRham theorem and Poincar\'{e} duality
\cite{ref:nakahara}. However, we will explicitly construct a set
of forms to show that they exist and possess all the desired
properties on the $N$-punctured genus $g$ torus.

Regarding the forms associated with the spin network punctures; it is convenient to arrange these punctures in the vicinity of a single toroidal segment, as show in the right diagram of figure \ref{fig:cycles}. We can define a set of coordinates $X,\,Y$ here such that the $X$ axis spans the poloidal direction of this particular toroidal hole and the $Y$ axis spans the toroidal direction.

\begin{figure}[ht]
\begin{center}
\includegraphics[bb=0 0 375 257, clip, scale=0.7, keepaspectratio=true]{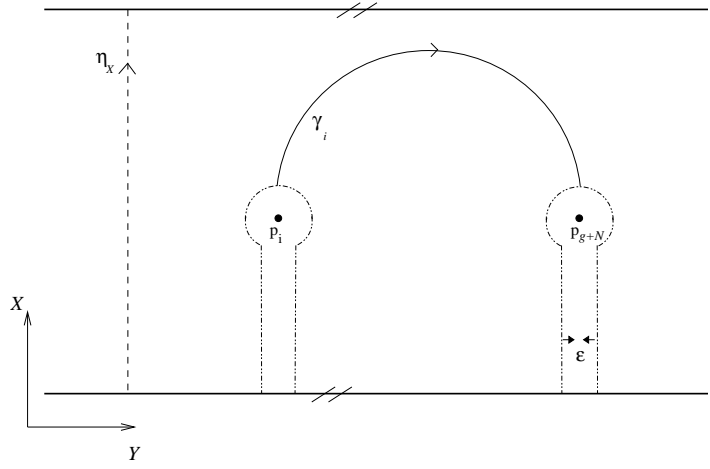}
\caption{{\small Branch cuts used to define the domain
$U^{\epsilon}$ in order to ensure single-valued forms. The
keyhole regions surrounded by the dash-dot lines are excised. We
consider the limit $\varepsilon \rightarrow 0$. The smoothing
functions, $s$, ensure single valuedness away from the punctures
and in the neighborhood of the identified lines.}}
\label{fig:branch}
\end{center}
\end{figure}

The domain for the $\beta$ forms is obtained from $T$ by
cutting out circles of radius $\varepsilon$ around $p_{i}$ and
$p_{N}$, and strips of width $\varepsilon$ as shown in figure
\ref{fig:branch}.

Let $\theta_{i}$ be the angle around $p_{i}$.
We measure the angle from a point some distance directly above
$p_{i}$. The $\gamma_{i}$ approaches $p_{i}$ at $\theta_{i}=0$
(this is a necessary part of the construction). Define
\begin{equation}
t_{i}:=s_{i}(X,\,Y)\,\theta_{i} - s_{g+N}(X,\,Y)\, \theta_{g+N}\,,
\end{equation}
where $s_{i}$ and $s_{g+N}$ are smoothing functions defined on all
of $g^{2}$.

Next, choose neighborhoods $V_{i}$ of $p_{i}$, $g+1\leq i \leq g+N$,
and $V_{X,\,Y}$ containing the local toroidal cycle which are all
disjoint. On $V_{i}$, choose $s_{i}=1$ and $s_{g+N}=0$, for $i \leq
N-1$. On $V_{g+N}$ choose $s_{i}=0$ and $s_{g+N}=1$. Finally, on
$V_{X,\,Y}$ choose both $s_{i}$ and $s_{g+N}$ to be zero. The
properties of the smoothing functions, along with the definition
of the domain $U^{\varepsilon}$ ensures the single-valuedness of
the $\beta$ forms. 

In $U^{\varepsilon}$ we define $\beta_{i}^{\varepsilon}$ to be
$\frac{1}{2\pi}dt_{i}$. Note that $-\pi < \theta_{i} < \pi$. For any curve
$\gamma$, define $\beta_{i}(\gamma)= \lim_{\varepsilon
\rightarrow 0} \beta^{\varepsilon}_{i}(\gamma)$. Then one can
check that $\beta_{i}(\eta_{j})= \delta_{ij}$ for $i,\,j \leq
g+N-1$, and $\beta_{i}(\eta_{X})=0$ along with
$\beta_{i}(\gamma_{Y})=0$.

Recall that $\gamma_{i}$ goes from $p_{i}$ to $p_{g+N}$, so
$\gamma_{i}(0)=p_{i}$ and $\gamma_{i}(1)=p_{g+N}$. For small
$\varepsilon$, $\gamma_{i}(\varepsilon)$ will be in $V_{i}$, and
$\hamma_{i}(1-\varepsilon)$ will be in $V_{g+N}$. In both cases,
$\gamma_{i}$ approaches the point from the vertical direction.
Thus,
\begin{align}
&\beta_{i}(\gamma_{i})=\lim_{\varepsilon\rightarrow 0}
\beta^{\varepsilon}_{i}(\gamma_{i})= \lim_{\varepsilon
\rightarrow 0} \int_{\gamma_{i}} \beta_{i}^{\varepsilon}
\nonumber \\
&= \frac{1}{2\pi} \left[\lim_{\varepsilon \rightarrow 0}\,
t_{i}(\gamma_{i}(1-\varepsilon)) -\lim_{\varepsilon \rightarrow
0}\, t_{i}(\gamma_{i}(\varepsilon))\right] =0 . \label{betadef}
\end{align}
In a similar way, $\beta_{i}(\gamma_{j})=0$ for $j\neq i$.

Therefore, the following forms, with some minor restrictions to be
discussed shortly, possess the desired properties (1 - 3) above:
\begin{subequations}
\begin{align}
 &\alpha_{j}=df_{j}(Y)\,,\;\;\;\;\;\;\;\; \alpha_{m}=df_{m}=f_{m}^{\prime}(Y_{m})\,dY_{m} \\
 &\beta_{j}=\lim_{\varepsilon \rightarrow 0} \beta^{\varepsilon}_{j}\,,
 \;\;\; \;\;\;\;\; \beta_{m}=dX_{m}\,.
\end{align}
\end{subequations}
Where the function $f$ is defined below.
Here, the prime denotes a partial derivative with respect to the
$Y_{m}$ coordinate, a coordinate spanning the toroidal direction on the $m$-th torus (recall from the construction that all spin network punctures are located on the $m=1$ torus). $X_{m}$ is the corresponding poloidal coordinate. 

\begin{figure}[ht]
\begin{center}
\includegraphics[bb=0 0 895 262, clip, scale=0.45, keepaspectratio=true]{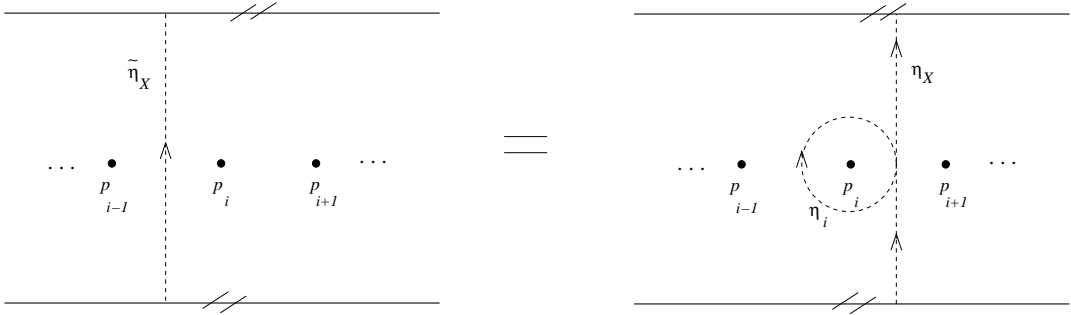}
\caption{{\small A schematic depicting the transformation in (\ref{eq:etaxtransf}) on the toroidal segment containing the punctures.}} \label{fig3}
\end{center}
\end{figure}

Note that the location of the cycle $\eta_{X}$ around the local poloidal direction is irrelevant in
the sense that, in the basis provided, a cycle circling the torus
in this direction can be transformed to another other cycle
circling the $X$ direction in an adjacent sector of the torus.
Schematically, we write this transformation as:
\begin{equation}
 \eta_{X} \rightarrow \tilde{\eta}_{X} = \eta_{i}^{-1}\eta_{X} \label{eq:etaxtransf}
\end{equation}
 (see figure \ref{fig3}). By repetition of the above transformation, the $\eta_{X}$
 cycle can be transferred to any sector of this particular toroidal segment. 

From property 3 we find that all punctures must be arranged on the
same $X=\mbox{constant}$ line  so, if $f_{{1}}$ were a one-to-one
function, we would be forced to choose $Y_{1}(p_{i}) = Y_{1}(p_{g+N})$,
which is not possible. We only require that
$f_{{1}}(1)-f_{{1}}(0)=1$, and we could have a function such as
\begin{align}
f_{1}(Y_{1})=&\sin\left[\left(2\pi N +\frac{\pi}{2}\right)Y_{1}\right], \;\;\;\; \mbox{for}\;\; m=1 \\
f_{m}(Y_{m})=& Y_{m}, \;\;\;\;\;\;\;\;\;\;\;\;\;\;\;\;\;\;\;\;\;\;\;\;\;\;\;\;\;\;\; \mbox{for}\;\; m > 1\label{eq:fY}
\end{align}
The $Y_{1}$ coordinate satisfies the condition
\begin{equation}
 Y_{1}(p_{j})-Y_{1}(p_{g+N}) = \frac{2\pi j}{2\pi N + \frac{\pi}{2}}
\end{equation}
for some $j$. The choice of $N$ is fixed in the definition of $f_{1}$.

Let us denote a generalized $U(1)$ connection as an expansion
\cite{ref:ashbaez}
\begin{equation}
 W=\sum_{i=g+1}^{g+N-1}\left(A_{i}\alpha_{i}+B_{i}\beta_{i}\right) +\sum_{m=1}^{g}A_{m}\alpha_{m}+B_{m}\beta_{m}=:
  \sum_{r=1}^{g+N-1} \left(A_{r}\alpha_{r}+B_{r}\beta_{r}\right). \label{eq:wexplicitXY}
\end{equation}
with $A_{r}$ and $B_{r}$ being elements of the $u(1)$ algebra. 
  Then we can show that $W$ is
flat, except at the punctures, and has singularities of the
standard type. Note that $d\alpha_{j}=ddf_{j}=0$. As well,
$d\alpha_{m}=ddf_{m}=0$ so that all the $\alpha$'s are flat.
Also, the $\alpha$ forms possess no singularities and all
singularities are contributed to $W$ by the $\beta$ forms. Now,
by construction, all singularities of $\beta_{j}$ are of the
standard form. Also, away from the punctures we have that
$d\beta_{j}=0$, so $\beta_{j}$ is flat there. With respect to
$\beta_{m}$, the condition $d\beta_{m}=ddX_{m}=0$ exists, so
$\beta_{m}$ is flat everywhere and possesses no singularities.
Thus, $\alpha_{m}$ and $\beta_{m}$ can be added to the sum as was
done in (\ref{eq:wexplicitXY}). Note that, although the $U(1)$
connection is associated with the gravitational spin network
punctures, one can easily extend the definition to incorporate
the toroidal holes, which also naturally have a $U(1)$ structure
associated with them. Therefore, this construction allows us to
transparently account for the toroidal holes in the counting of
the degrees of freedom.

As mentioned earlier, in the case of the genus $g$ torus, the analog of
(\ref{eq:projconst}) is taken care of by (\ref{eq:torusrelg}). Note
that this condition, like (\ref{eq:projconst}), is related to the topology of the surface.

We next define
\begin{subequations}
\begin{align}
 j_{I}(W)=W(\gamma_{I}), \label{eq:gh1} \\
 h_{I}(W)=W(\eta_{I}), \label{eq:gh2}
\end{align}
\end{subequations}
which are holonomies and elements of $U(1)$ and where
the subscript $I=1,\,2,\,...,\, g,\,g+1,\,...,\,g+N-1$. This definition
is as in \cite{ref:ashbaez} save for the fact that there now
exist effective holonomies from the degrees of freedom due to the toroidal cycles.
 A map $\Phi$ can be defined that maps the phase-space
 to a $2(g+N-1)$ dimensional torus $\left(U(1)\times U(1)\right)^{g+N-1}$. $\Phi$ is defined as
\begin{equation}
 \Phi\left([W]\right)=\left\{j_{1}(W),\,h_{1}(W),\,...,\,j_{g}(W),\,h_{g}(W),\, ...,\, j_{g+N-1}(W),\,h_{g+N-1}(W)\right\},
\end{equation}
an ordered $2(g+N-1)$-tuple of the holonomies. Let
$(A_{1},\,B_{1},\,...,\,A_{g+N-1},\,B_{g+N-1})$ be a $2(g+N-1)$-tuple in
$\left(U(1)\times U(1)\right)^{g+N-1}$. Let $W$ be defined as in
(\ref{eq:wexplicitXY}). Then,
\begin{align}
 \Phi\left([W]\right)=&\left\{j_{1}(W),\,h_{1}(W),\,...,\,j_{g}(W),\,h_{g}(W),\, ...,\, j_{g+N-1}(W),\,h_{g+N-1}(W)\right\}
 \nonumber \\
 =&\left\{W(\gamma_{1}),\, W(\eta_{1}),\,...,\, W(\gamma_{g+N-1}),\,
 W(\eta_{g+N-1})\right\} = \left\{A_{1},\,B_{1},\,...,\,
 A_{g+N-1},\,B_{g+N-1}\right\},
\end{align}
using properties 1 and 2 of the forms. Thus $\Phi$ is onto.

Next, note that
\begin{equation}
 \Phi(W)=\Phi(W^{\prime})\;\;\; \mbox{iff}\;\;\; j_{I}(W)=j_{I}(W^{\prime})\;\;\;
 \mbox{and}\;\;\; h_{I}(W)=h_{I}(W^{\prime}). \label{eq:phicond}
\end{equation}
From (\ref{eq:phicond}), $W(\gamma_{I})=W^{\prime}(\gamma_{I})$ and $W(\eta_{I})=W^{\prime}(\eta_{I})$. Thus, $W$ and $W^{\prime}$ agree on $\left\{\gamma_{1},\,...,\, \gamma_{g+N-1}\right\}$ and on $\left\{\eta_{1},\,...,\, \eta_{g+N-1}\right\}$. Hence $W$ and $W^{\prime}$ agree on all the paths in (\ref{eq:torbasis}). Since this is a basis for all paths based at $p_{g+N}$, $W$ and $W^{\prime}$ will agree on all paths. Thus $W$ and $W^{\prime}$ are the same.
This is important to ensure that we are counting all states, and are not overcounting, for the black hole entropy calculation.

From the boundary term in (\ref{eq:gravsymp}), the symplectic
structure on the phase-space is given by
\begin{equation}
 \omega\left(\delta W,\,\delta W^{\prime}\right)
 =\frac{k}{2\pi}\oint_{T} \delta W \wedge \delta W^{\prime}\,, \label{eq:symstruct2}
\end{equation}
where $\delta W$ and $\delta W^{\prime}$ belong to the space of generalized $U(1)$ 
connections. The properties of the one-forms yield:
\begin{align}
 \oint_{T}\delta W \wedge \delta W^{\prime}=& \oint_{T} \sum_{n=1}^{g+N-1} \left(\delta A_{n}\alpha_{n} +\delta B_{n} \beta_{n}\right) \wedge \sum_{m=1}^{g+N-1} \left(\delta A^{\prime}_{m}\alpha_{m} +\delta B^{\prime}_{m} \beta_{m}\right) \nonumber \\
 =&\oint_{T} \sum_{m,n}\left[\delta A_{n} \delta A^{\prime}_{m}\, \alpha_{n}\wedge \alpha_{m} + \delta A_{n} \delta B_{m}^{\prime} \,\alpha_{n} \wedge \beta_{m} + \delta B_{n} \delta A^{\prime}_{m} \,\beta_{n} \wedge \alpha_{m} \right. \nonumber \\
 & \left.  +\delta B_{n} \delta B^{\prime}_{m}\, \beta_{n}\wedge \beta_{m} \right] \nonumber \\
 =&\sum_{m,n}\left[ 0 + \delta A_{n} \delta B_{m}^{\prime}\, \delta_{nm} + \delta B_{n} \delta A_{m}^{\prime} \left(-\delta_{nm}\right) +0 \right] \nonumber \\
 =& \sum_{n} \left[ \delta A_{n} \delta B^{\prime}_{n} - \delta B_{n} \delta A^{\prime}_{n}\right]. \label{eq:sympstruct}
\end{align}
Here, the third property of the forms has been used. This is the explicit symplectic structure in the surface  phase-space and illustrates that this phase-space is diffeomorphic to a $\left(U(1)\times U(1)\right)^{g+N-1}$ torus.

\subsection{The $g=1$ cases}
In the case of a genus $1$ surface, the metric can be written as:
\begin{equation}
 ds^{2}=-\left(\alpha^{2} \rho^{2} -\frac{2M}{\rho}\right)\,dt^{2}+ \frac{d\rho^{2}}{\left(\alpha^{2} \rho^{2} -\frac{2M}{\rho}\right)} +\rho^{2}\,d\theta^{2} +\rho^{2}\theta^{2}\,d\phi^{2}\:. \label{eq:g1surfmetric}
\end{equation}
This case is special in the sense that the fiducial $U(1)$ connection yields a constant connection on the horizon which is therefore flat. This gives zero for the boundary term in (\ref{eq:deltavar}). $F$ and $\Sigma$ in this case decouple, as can be seen from the fact that the first term in (\ref{eq:pullback}) yields a zero coefficient. Recalling that the boundary term originally arose from the coupling of $\Sigma$ and $A$, this is not surprising. 

However, black hole entropy has been calculated for $g=1$ black holes utilizing non loop quantum gravity techniques \cite{ref:vanzo}, \cite{ref:mans}. One can extend the loop quantum gravity method for $g\neq 1$ to $g=1$ by noting that a term proportional to the integral in (\ref{eq:u1form}) may be added to the classical action in the $g=1$ case without affecting the classical equations of motion (as we are adding zero). We will show here that a set of non-trivial connections may be constructed on the way to quantization when one punctures the torus, utilizing exactly the same methods as above. This will be shown to yield an entropy of $A/4$ with \emph{no} logarithmic correction, which is exactly consistent with findings in the literature using non loop quantum gravity approaches \cite{ref:vanzo} \cite{ref:liko}. 

Since the construction here is very similar to the $g>1$ case, we will present it in brief.

\subsubsection{The toroidal horizon}

\begin{figure}[ht]
\begin{center}
\includegraphics[bb=0 0 640 261, clip, scale=0.5, keepaspectratio=true]{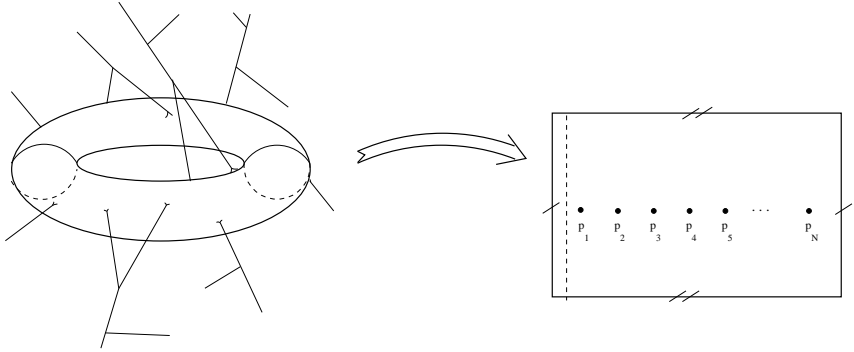}
\caption{{\small The toroidal horizon is punctured by the
gravitational spin network, giving it geometry and area. To
construct the phase-space we consider the identified plane with $N$
punctures.}} \label{fig1}
\end{center}
\end{figure}

As with the $g>1$ scenario, the horizon is mapped to a polygon, now as in figure \ref{fig1}.

In this case, a basis for all paths based at $p_{N}$ is
\begin{equation}
 \left\{\gamma_{1}^{-1}\eta_{1}\gamma_{1},\, ...,\, \gamma_{N-1}^{-1}\eta_{N-1}\gamma_{N-1},\, \gamma^{-1}_{Y}\eta_{X}\gamma_{Y}\right\}, \label{eq:torbasis}
\end{equation}
which will break down to
$\left\{\eta_{1},\,...,\,\eta_{N-1},\,\eta_{X}\right\}$ and
$\left\{\gamma_{1},\,...,\,\gamma_{N-1},\,\gamma_{Y}\right\}$. . Note that, like the other cases, $\eta_{N}$ again
is not used as it can be expressed using the relation
\begin{equation}
\eta_{1}\cdot\eta_{2}\cdot ... \cdot \eta_{N}=\gamma_{Y}\eta_{X}\gamma_{Y}^{-1}\eta_{X}^{-1}  \label{eq:torusrel}
\end{equation}
The
result is $2N$ degrees of freedom for the $g=1$ torus. 

Before continuing, it is interesting to note that this result may also be obtained directly from the following argument relating $g=2$ to $g=1$: 
Let $T$ be a $g=1$ torus with $N$ punctures, and let $T^{\prime}$ be another $g=1$ torus with
$N$ punctures. Imagine identifying one puncture of $T$ with one puncture of $T^{\prime}$.
The result will be a torus of genus 2, with $N_{tot}=2N-2$ punctures. Now since $g>1$ we
can use the argument that the phase space will have dimension $2[g+N_{tot}-1]
= 2[2N-1]$. It is reasonable to assume that each of $T$ and $T^{\prime}$ contribute half of this
dimension, or $2N-1$. Note that we have 
lost 2 degrees of freedom doing the identification, and hence each individual torus, if considered separately,
should acquire one more degree of freedom if the argument is applied backwards (i.e. the two tori are separated), yielding $2N$.

\begin{figure}[ht]
\begin{center}
\includegraphics[bb=0 0 364 248, clip, scale=0.6, keepaspectratio=true]{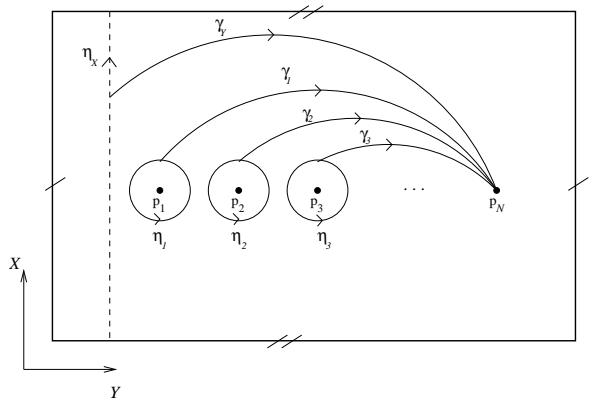}
\caption{{\small Open chains ($\gamma$'s) and cycles ($\eta$'s)
used to define forms for the symplectic structure of the
phase-space of the toroidal horizon.}} \label{fig:2}
\end{center}
\end{figure}

The forms $\alpha$ and $\beta$ must possess similar properties as discussed in the previous section. As well, similar to the previous case, chains in the direction of the torus angles
($X$ and $Y$ in figure (\ref{fig:2})) are taken as conjugates of
each other.
The construction now follows a similar argument as in the $g>1$ scenario (and is in fact, simpler). We calculate explicitly that:
\begin{subequations}
\begin{align}
 &\alpha_{j}=df_{j}\,,\;\;\;\;\;\;\;\; \alpha_{Y}=df_{Y}=f_{Y}^{\prime}(Y)\,dY \\
 &\beta_{j}=\lim_{\varepsilon \rightarrow 0} \beta^{\varepsilon}_{j}\,,
 \;\;\; \beta_{X}=dX\,,
\end{align}
\end{subequations}
with $f$ having a similar definition to the function in the $g>1$ case. 

Let us denote a generalized $U(1)$ connection as an expansion
\begin{equation}
 W=\sum_{i=1}^{N-1}\left(A_{i}\alpha_{i}+B_{i}\beta_{i}\right) +A_{Y}\alpha_{Y}+B_{X}\beta_{X}=:
  \sum_{r=1}^{N} \left(A_{r}\alpha_{r}+B_{r}\beta_{r}\right). \label{eq:wexplicitXYtor}
\end{equation}
Note that even though we started with a surface with constant connection, we can define a set of connections \emph{flat everywhere except at the punctures} when we go over to a picture of a punctured surface.

Note that for the $g=1$ torus, the analog of
(\ref{eq:projconst}) is taken care of by (\ref{eq:torusrel}). 

The symplectic
structure on the phase-space is now given by
\begin{equation}
 \oint_{T^{2}}\delta W \wedge \delta W^{\prime}= \sum_{n=1}^{N} \left[ \delta A_{n} \delta B^{\prime}_{n} - \delta B_{n} \delta A^{\prime}_{n}\right]. \label{eq:sympstructtor}
\end{equation}
Here, the third property of the forms has been used. This is the explicit symplectic structure in the surface  phase-space and illustrates that this phase-space is diffeomorphic to a $\left(U(1)\times U(1)\right)^{N}$ torus.

\subsubsection{The cylindrical horizon}
Here we consider the cylindrical horizon. Due to the non-compact nature of this surface we are interested in the entropy per unit  area and therefore divide the cylinder into segments with $N$ punctures each. We will find the entropy and area for each segment, and thus the relationship between them. Let $C$ denote one of the segments, and $P_{N}=\left\{p_{1},\,...,\,p_{N}\right\}$ be the $N$ punctures in this segment. Then $dim H^{1}(C-P_{N})=n+1$, the same as the torus. The construction of the phase space is different, however.

\begin{figure}[ht]
\begin{center}
\includegraphics[bb=0 0 901 304, clip, scale=0.45, keepaspectratio=true]{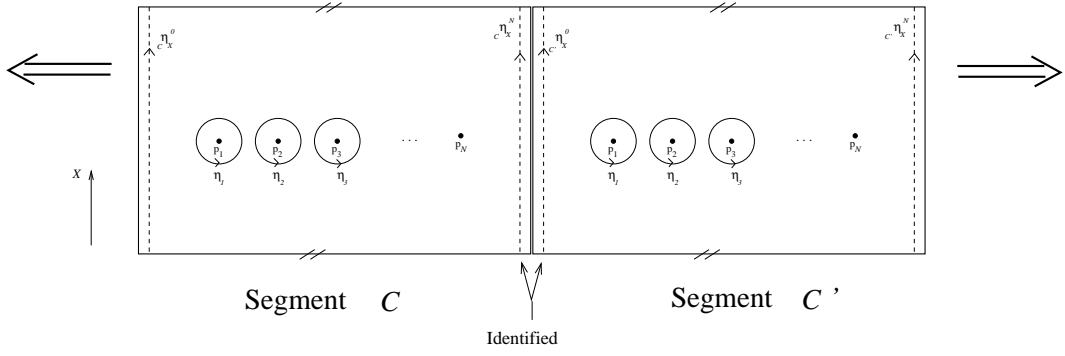}
\caption{{\small The cycles ($\eta$'s)
used to define forms for the symplectic structure of the
phase-space for a segment of the cylindrical horizon. $C$ and $C^{\prime}$ represent two adjacent segments of the cylinder. Note that it appears that there are two independent cycles per segment associated with the axial direction of the cylinder in each section (eg. $_{C}\eta^{0}_{X}$ and $_{C}\eta^{N}_{X}$). However, one of these cycles ($_{C}\eta^{N}_{X}$)is actually associated with the adjacent segment $C^{\prime}$ and is equivalent to $_{C^{\prime}}\eta^{0}_{X}$ and therefore there are actually no extra degrees of freedom.}} \label{fig:cyl}
\end{center}
\end{figure}

Below, in figure \ref{fig:cyl}, is the corresponding diagram for $C$ and  an adjacent section $C^{\prime}$. The top is identified with the bottom, but the left is now \emph{not} identified with the right.

Let $\eta_{1},\,...,\,\eta_{N}$ be closed loops around the punctures. Let $\eta^{0}_{X}$ be a loop going around in the $X$ direction which is to the left of $\eta_{1}$ and let $\eta^{N}_{X}$  be a similar loop which is to the right of $\eta_{N}$. Construct $\gamma_{1},\,...,\, \gamma_{N-1}$ in the usual way. For $\gamma_{N}$, we draw a curve from $p_{N}$ to $\eta^{0}_{X}$ and for $\gamma_{N+1}$, we draw a curve from $p_{N}$ to $\eta^{N}_{X}$. Next, let $\hamma:=\gamma^{-1}_{N+1}\gamma_{N}$. Then $\gamma$ furnishes a curve connecting $\eta^{N}_{X}$ $\eta^{0}_{X}$. The fundamental loop relation can be written as 
\begin{equation}
 \eta_{1}\cdot\eta_{2}\cdot ... \cdot \eta_{N}= \eta^{N}_{X} \gamma (\eta^{0}_{X})^{-1}\gamma^{-1}. \label{eq:cylrel}
\end{equation}
A basis for the loops based at $p_{N}$ is $\left\{\eta_{1},\,...,\,\eta_{N-1};\,\eta^{0}_{X},\,\eta^{N}_{X}\right\}\cup \left\{\gamma_{1},\,...,\,\gamma_{N};\,\gamma_{N+1}\right\}$ because we can express $\eta_{N}$ by using (\ref{eq:cylrel}). Care must be taken not to overcount, as this segment of $C$ will be joined on both sides to similar segments. For example, if $C^{\prime}$ is the segment to the right of $C$, then the loop $\eta_{X}^{N}$  of $C$ is the same as $\eta_{X}^{0}$ of $C^{\prime}$. We construct the phase-space for each segment utilizing only the left loop of the two. Then the phase-space will be $\left[U(1)\times U(1)\right]^{N}$. Thus we are reduced to the same situation as the torus with $N$ punctures and we can make use of the results for the torus. The details of the construction of the $W$'s is omitted here as arguments are similar to those of the torus, with minor technical modifications, and do not provide any new insight into the phase-space structure.

\section{Relation to black hole entropy}

Having constructed the phase-space, we can now count its
dimension to determine the entropy associated with the horizon.
As described in the previous sections, we have a phase space with
a $\left(U(1)\times U(1)\right)^{N+g-1}$ structure, where $N$ is the number of punctures. This phase-space is therefore diffeomorphic to
a $2(N+g-1)$ dimensional torus.

The entropy calculation follows much the same lines as in [11] - in 
fact, our calculation exploits the fact that our situation can be mapped 
onto the original situation considered in \cite{ref:ashbaez}. An important fact to 
note about this mapping is that the dimensionality of the surface 
Hilbert space is different in the two cases and the end result will 
therefore differ as well. Let us briefly sketch
the derivation; for details the reader is referred to
\cite{ref:ashbaez}, \cite{ref:entrev6} and \cite{ref:domlew}.

The `prequantization condition' involving $k$, the level of the Chern-Simons
theory is discussed in \cite{ref:ashbaez}. Eq. (15) of that paper can be generalized to

\begin{equation*}
k=\frac{a_{0}}{4\pi \gamma \ell _{P}^{2}\chi }
\end{equation*}
where $\chi $ is the Euler characteristic of the horizon $T$, and $a_{0}$ is
the area. The next step is to compare the surface eigenvalues with the
bulk eigenvalues. We find the condition for the spectra to coincide is 
\begin{equation*}
a_{i}=\frac{2m_{i}}{g-1}\text{ }\mbox{mod}k
\end{equation*}
Since the \ $m$'s are half-integers and the $a_{i}$'s are integers, we see that
the only allowed values of $g$ are $0$ and $2$. The only difference between
these cases is a change of sign in the above equation, and the counting of
states is closely connected. Our consideration of higher genus has revealed
that finding a quantization scheme to cover all values of $g$ is an open
question. We will show how the quantum states can be counted for $g=2$.

Taking $g=2$, we can define 'permissible' lists as in \cite{ref:ashbaez}. Given an ordered
list of positive half-integers $j$=($j_{1}$,...,$j_{n}$), let $A(j)$ be the
corresponding eigenvalue of the area operator:

\begin{equation}
A(j)=8\pi \gamma \ell _{P}^{2}\sum_{i}\sqrt{j_{i}(j_{i}+1)} \label{eq:area_eigen}
\end{equation}

We say the list $j$ is 'permissible' if it satisfies

\begin{equation*}
a_{0}-\delta \leq A(j)\leq a_{0}+\delta .
\end{equation*}

Given a list of half-integers $(m_{1},...,m_{n}),$ we say it is
'permissible' if for some permissible list of positive half-integers $j$=($%
j_{1}$,...,$j_{n}$), we have $-j_{i}\leq m_{i}\leq j_{i}.$ Finally, given a
list $(a_{1},...,a_{n})$ of elements of $Z_{k}$, we say that it is
'permissible' if $a_{i}=-2m_{i}\mbox{mod}k$ for some permissible list of
half-integers $(m_{1},...,m_{n}).$ We do not require the constraint in eq.
(\ref{eq:projconst}) as it is eliminated by (\ref{eq:torusrelg}). Let $Q$ be the set of permissible lists $a$.

Let $P$ be a finite set of points on the horizon, and let $H_{T}^{P}$ be the
surface Hilbert space. As in \cite{ref:ashbaez}, a basis for this space will be $\psi _{a}$%
, where $a$ ranges over the set $Q.$ To determine the entropy, we want to
count the number of elements in this set, $N(Q).$ In our case, $n$ is $2+N,$
and the permissible lists will be of the form $%
(a_{1},a_{2,}a_{2+1},...,a_{2+N})$, where the first 2 values are related to
the topology, and the other $N$ values are related to the punctures.

Let us consider the number of states that arise from the last $N$ variables.
We can relate this number to the case of a spherical boundary, which was
considered in \cite{ref:entrev6}. There the states were counted by a computer program, simply counting the number of states which yield a surface of fixed area,
avoiding the approximation problems introduced by analytic approaches. They
found the Immirzi parameter that yields the asymptotic linear relation
compatible with the Bekenstein-Hawking entropy is close to $\gamma =0.274$.
By doing the counting both with and without the constraint, they found that
the constraint (\ref{eq:projconst}) is exactly responsible for the logarthmic correction. In
our case, the constraint is absent, and hence the logarithmic term is
absent. The entropy contribution from the last $N$ variables is thus $A/4$.

We also have an entropy contribution from the first 2 variables. Because of
the constraint (\ref{eq:torusrelg}), we can only assign values to one of them. Since this
variable does not enter into the area formula, it may possess any value
between 1 and k. The number of such states is $k$, so the total entropy, $S$%
, will be 
\begin{equation*}
S=A/4+\ln k=A/4+[\ln A-\ln 4\pi \gamma ],
\end{equation*}
where $k$ is the Chern-Simons level as previously defined.\qquad

Although we do not have a strict quantization scheme for genus higher than 2, we
can make some speculations about the entropy given the topological counting of degrees of freedom done here. One can extend the above result and conjecture that the entropy will be

\begin{equation}
S=A/4+(g-1)[\ln A-\ln (4\pi \gamma (g-1))], \label{eq:theentropy}
\end{equation}
since the number of states due to the topological variables is $k^{g-1}$. We conclude that $S$ has a dependence on genus in the subdominant term. These results are in qualitative agreement with other studies which find $g$-dependence corrections using non-loop quantum gravity techniques \cite{ref:vanzo}, \cite{ref:mans} (with a similar form for the genus dependent coefficient in the sub-dominant term) and leads us to believe that this is correct extension of the quantization condition here. As we comment next, (\ref{eq:theentropy}) is also in agreement with $g=1$ calculations in the literature which utilise non-loop quantum gravity techniques.

In the case of the genus-1 torus (and the cylinder), the logarithmic correction term is not present. This is due to the fact that the condition (\ref{eq:etarel}) is
replaced by (\ref{eq:torusrel}) in this case, which removes the degeneracy in counting. This result for the $g=1$ entropy is consistent with the result found by Vanzo \cite{ref:vanzo} utilizing a Euclidean quantum field theoretic approach. This is also the same result obtained in \cite{ref:liko} where a Gauss-Bonnet term was added to the Einstein-Hilbert action and a thermodynamic approach was utilised. A discussion of the significance and impact of adding an arbitrary
coupling constant of the GB term in asymptotically AdS space-times may be found in the work of Olea \cite{ref:olea}.

For the case of the cylindrical horizon, Let $Q_{C}$ the set of quantum states that we attribute to $C$. Then, $S_{C}=\ln [N(Q_{C})]=A/4$, where $A$ is now the area arising from the punctures of the segment $C$ (again we have utilized the careful counting result of \cite{ref:entrev6} without the spherical constraint). In a similar way, if $Q_{C^{\prime}}$ is the set of quantum states attributable to $C^{\prime}$, then $S_{C^{\prime}}=\ln [N(Q_{C^{\prime}})]=A^{\prime}/4$. Now if we consider the quantum states corresponding to $C+C^{\prime}$, we see that $[N(Q_{C+C^{\prime}})]=[N(Q_{C})][N(Q_{C^{\prime}})]$, so $S_{C+C^{\prime}}=\ln [N(Q_{C})]+\ln [N(Q_{C^{\prime}})]=\frac{A+A^{\prime}}{4}$. Thus the same relation holds for any length of the cylinder and the entropy per unit segment is given by $S_{cyl}=A/4$, the same as the $g=1$ torus.

It is interesting to note that the formula for the entropy-area relationship
does not depend on the value of $N$. Thus the formula must be the same for
two segments as it is for one segment, and hence there cannot be any
logarithmic term.

We note that the entropy calculation however cannot be extended to the $g=0$ case, due to the fact that the genus cycles in this construction play a crucial role and the $g=0$ scenario is qualitatively different from the others.

It is perhaps not surprising that the entropy associated with
higher genus horizons is larger than a spherical horizon at the
first-order correction level. The toroidal cycles introduce
degrees of freedom to the classical phase space that are not
present for $S^{2}$ topology. Therefore, it is expected that the
entropy be slightly larger for these cases.

\section{Concluding remarks}
In this paper we constructed the phase-space corresponding to an isolated horizon with $g \geq 1$ and cylindrical topology in the context of loop quantum gravity. We have included the toroidal and cylindrical  cycles as degrees of freedom and found that these degrees of freedom couple to the gravitational spin network degrees of freedom via the topological conditions (\ref{eq:torusrelg}) and (\ref{eq:cylrel}). The entropy of the horizon can be calculated by taking the logarithm of the number of surface states which yield a surface area equal to the classical area, $A$, of the horizon. This produces the leading order $A/4$ term provided the Immirzi parameter, $\gamma$, is set to the value solving the relation $\sum_j (2j + 1) \exp (2\pi \gamma j(j + 1) ) = 1$, which is the same value found in other calculations involving spherical horizons. The logarithmic correction term, which has been calculated to be $-\frac{1}{2}\ln A$ in the case of spherical horizons \cite{ref:entrev6}, is modified in this calculation due to the presence of the toroidal and cylindrical cycles. We attribute this to the the larger number of degrees of freedom introduced by these cycles which are not present in the spherical case. One can most easily see how these cycles contribute to the entropy by considering the hypothetical case of horizons with a single puncture. The sum for the area in all cases will be based on that puncture, but the
entropy count will be different. The toroidal and cylindrical cycles allow the puncture to
contribute to the entropy, but in the spherical case there is no
corresponding contribution, since we only have $\eta_1$, and that is trivial.

\section*{Acknowledgments}
We would like to thank A. Ashtekar, A. Corichi, K. Krasnov, S.
Major and D. Witt for their useful input and helpful discussions.
JB would like to acknowledge partial support from The Blanceflor
Foundation and the hospitality of the Perimeter Institute, where some of this work was carried out.

\linespread{0.6}
\bibliographystyle{unsrt}

\begin{thebibliography}{10}
{\small

\bibitem{ref:bek}
J.~D. Bekenstein,
\newblock {\em Phys. Rev.} {\bf D7} (1973) 2333.

\bibitem{ref:entreview}
P. Majumdar,
\newblock {\em gr-qc/9807045}.

\bibitem{ref:entreview2}
R.~M. Wald,
\newblock {\em Living Rev. Relativity} {\bf 4}  (2001)  6.

\bibitem{ref:thiem}
T. Thiemann,
\newblock {\em gr-qc/0110034}.

\bibitem{ref:rovbook}
C. Rovelli,
\newblock {\em Quantum Gravity} (Cambridge University Press, Cambridge, 2004).

\bibitem{ref:rev1}
A. Perez,
\newblock {\em gr-qc/0409061}

\bibitem{ref:rev2}
A. Ashtekar and J. Lewandowski,
\newblock {\em Class. Quant. Grav.} {\bf 21} (2004) R53.

\bibitem{ref:bojo}
M. Bojowald,
\newblock In {\em Trends in Quantum Gravity Research} (Nova Science Pub., New York, 2006).

\bibitem{ref:modesto}
L. Modesto,
\newblock {\em Class. Quant. Grav.} {\bf 23} (2006) 5587.

\bibitem{ref:entrev1}
C. Rovelli,
\newblock {\em Phys. Rev. Let.} {\bf 77} (1996) 3288.

\bibitem{ref:ashbaez}
A. Ashtekar, J.~C. Baez and K. Krasnov,
\newblock {\em Adv. Theor. Math. Phys.} {\bf 4} (2000) 1.

\bibitem{ref:daskaul}
S. Das, R.~K. Kaul and P. Majumdar,
\newblock {\em Phys. Rev.} {\bf D63} (2001) 044019.

\bibitem{ref:chatmaj}
A. Chatterjee and P. Majumdar,
\newblock {\em gr-qc/0303030v1}

\bibitem{ref:domlew}
M. Domagala and J. Lewandowski,
\newblock {\em Class. Quant. Grav.} {\bf 21} (2004) 5233.

\bibitem{ref:entrev2}
K.~A. Meissner,
\newblock {\em Class. Quant. Grav.} {\bf 21} (2004) 5245.

\bibitem{ref:entrev3}
A. Ghosh and P. Mitra,
\newblock {\em Phys. Let.} {\bf B616} (2005) 114.

\bibitem{ref:entrev3_5}
T. Tamaki and H. Nomura,
\newblock {\em Phys. Rev.} {\bf D72} (2005) 107501.

\bibitem{ref:dms}
O. Dreyer, F. Markopoulou and L. Smolin,
\newblock {\em Nucl.Phys.} {\bf B744} (2006) 1.

\bibitem{ref:entrev4}
A. Ghosh and P. Mitra,
\newblock {\em Indian J. Phys.} {\bf 80} (2006) 867.

\bibitem{ref:entrev5}
M.~H. Ansari,
\newblock {\em gr-qc/0603121}

\bibitem{ref:entrev6}
A. Corichi, J. D\'{i}az-Polo and E. Fern\'{a}ndez-Borja,
\newblock {\em Class. Quant. Grav.} {\bf 24} (2007) 243.

\bibitem{ref:immirz1}
K.~A. Meissner,
\newblock {\em Class. Quant. Grav.} {\bf 21} (2004) 5245.

\bibitem{ref:jacobson}
T. Jacobson,
\newblock {\em 	Class. Quant. Grav.} {\bf 24} (2007) 4875.

\bibitem{ref:kras}
K. Krasnov,
\newblock {\em Phys. Rev.} {\bf D55} (1997) 3505.

\bibitem{ref:countambig}
T. Tamaki,
\newblock {\em Class. Quant. Grav.} {\bf 24} (2007) 3837.

\bibitem{ref:rg1}
R.~G. Cai and Y.~Z. Zhang,
\newblock  {\em Phys. Rev.} {\bf D54} (1996) 4891.

\bibitem{ref:tor1}
J.~P.~S. Lemos and V.~T. Zanchin,
\newblock {\em Phys. Rev.} {\bf D54} (1996) 3840.

\bibitem{ref:tor2}
W.~L. Smith and R.~B. Mann,
\newblock {\em Phys. Rev.} {\bf D56} (1997) 4942.

\bibitem{ref:vanzo}
L. Vanzo,
\newblock {\em Phys.Rev.} {\bf D56} (1997) 6475.

\bibitem{ref:tor3}
J.~P.~S. Lemos,
\newblock {\em Phys. Rev.} {\bf D57} (1998) 4600.

\bibitem{ref:rg2}
R.~G. Cai, J.~Y. Ji and K.~S. Soh,
\newblock {\em Phys. Rev.} {\bf D57} (1998) 6547.

\bibitem{ref:rg3}
R.~G. Cai and K.~S. Soh,
\newblock {\em Phys. Rev.} {\bf D59} (1999) 044013.

\bibitem{ref:tor4}
S. Surya, K. Schleich and D. Witt,
\newblock {\em Phys.Rev.Let.} {\bf 86} (2001) 5231.

\bibitem{ref:liko}
T. Liko,
\newblock {\em 	arXiv:0705.1518v2} [gr-qc].

\bibitem{ref:mena}
F.~C. Mena, J. Nat\'{a}rio and P. Tod,
\newblock {\em 	arXiv:0707.2519} [gr-qc]

\bibitem{ref:ashcor}
A. Ashtekar, A. Corichi and K. Krasnov,
\newblock {\em Adv. Theor. Math. Phys.} {\bf 3} (2000) 419.

\bibitem{ref:thermo}
D.~R. Brill, J. Louko and P. Peld\'{a}n,
\newblock {\em Phys. Rev.} {\bf D56} (1997) 3600.

\bibitem{ref:ashlambda}
A. Ashtekar, T. Pawlowski and C. Van~Den~Broeck, ``Mechanics of
higher-dimensional black holes in asymptotically anti-de Sitter spacetimes''.
\newblock {\em in preparation}

\bibitem{ref:ashfair}
A. Ashtekar, S. Fairhurst and B. Krishnan,
\newblock {\em Phys. Rev.} {\bf D62} (2000) 104025.

\bibitem{ref:ashengle}
A. Ashtekar, J. Engle, T. Pawlowski and C. Van~Den~Broeck,
\newblock {\em Class. Quant. Grav.} {\bf 21} (2004) 2549.

\bibitem{ref:ashengle2}
A. Ashtekar, J. Engle and C. Van~Den~Broeck,
\newblock {\em Class. Quant. Grav.} {\bf 22} (2005) L27.

\bibitem{ref:abf}
A. Ashtekar, C. Beetle, S. Fairhurst, 
\newblock {\em Class. Quant. Grav.} {\bf 17} (2000) 253.

\bibitem{ref:jerzlewandhoriz}
J. Lewandowski,
\newblock {\em Class. Quant. Grav.} {\bf 17} (2000) L53.

\bibitem{ref:chandra}
S. Chandrasekhar,
\newblock {\em The Mathematical Theory of Black Holes}, (Oxford University Press, Oxford,1992).

\bibitem{ref:govind}
T.~R. Govindarajan, R.~K. Kaul and V. Suneeta,
\newblock {\em Class. Quant. Grav.} {\bf 18} (2001) 2877.

\bibitem{ref:mans}
R.~B. Mann and S.~N. Solodukhin,
\newblock {\em Nucl.Phys.} {\bf B523} (1998) 293.

\bibitem{ref:lowdim1}
S. Carlip,
\newblock {\em gr-qc/9305020}.

\bibitem{ref:lowdim2}
S. Carlip,
\newblock {\em Quantum gravity in 2+1 dimensions}, (Cambridge University Press, Cambridge, 1998).

\bibitem{ref:nakahara}
M. Nakahara,
\newblock {\em Geometry, Topology and Physics}, 2nd ed. (IOP, London, 2003).

\bibitem{ref:ashlewgeo}
A. Ashtekar and J. Lewandowski, 
\newblock {\em Class. Quant. Grav.} {\bf 14} (1997) A55.

\bibitem{ref:olea}
R. Olea, {\em J.H. E. P.} {\bf 0506} (2005) 023.

}

\end{thebibliography}

\end{document}